\documentclass[]{spie}  

\usepackage[]{graphicx}
\title{The LAUE project for broadband gamma-ray focusing lenses} 
\author{E.~Virgilli\supit{a}, F.~Frontera\supit{a,b}, V.~Valsan\supit{a,c}, V.~Liccardo\supit{a,c}, E.~Caroli\supit{b}, 
J.B.~Stephen\supit{b}, F.~Cassese\supit{d}, L.~Recanatesi\supit{d}, M.~Pecora\supit{e}, S.~Mottini\supit{f}, 
P.~Attin\'a\supit{a,f}, B.~Negri\supit{g}
\skiplinehalf
\supit{a} \small\textit{Physics Department, University of Ferrara - Italy};\\
\supit{b} \small\textit{IASF-INAF via P.Gobetti, Bologna - Italy};\\
\supit{c} \small\textit{Universit\'e de Nice Sophia-Antipolis, Parc Valrose, 06108 Nice Cedex 2, France};\\
\supit{d} \small\textit{DTM, Modena, Via Tacito, I-41100 Modena, Italy};\\
\supit{e} \small\textit{Thales Alenia Space-Italy, Milan, Italy};\\
\supit{f} \small\textit{Thales Alenia Space-Italy, Turin, Italy};\\
\supit{g} \small\textit{ASI, Agenzia Spaziale Italiana, Viale Liegi 26, I-00198 Roma, Italy}.
}

\authorinfo{E-mail to: frontera@fe.infn.it}

\begin{document} 
\maketitle

\begin{abstract}
We present the LAUE project devoted to develop an advanced  technology for building a high 
focal length Laue lens for soft gamma--ray astronomy (80-600 keV). The final goal is to develop
a focusing optics that can improve the current sensitivity in the above energy band by 2 
orders of magnitude.
\end{abstract}


\section{INTRODUCTION}
\label{sec:intro}

The astrophysical importance of the X--ray broad band (0.1--200~keV and beyond) has been 
demonstrated by missions like 
$Beppo$SAX, 
XTE, 
and INTEGRAL.
This band has been shown to be crucial to 
get a complete physical description of the astrophysical sources, like to establishing the 
geometry of the systems, the physical phenomena occurring in the emission 
region and the radiation production mechanisms. Furthermore, it enables us to distinguish 
the contribution of 
thermal emission phenomena from those due to the presence of high energy non thermal plasma 
and/or magnetic fields.
The focusing telescopes in the soft gamma--ray band are nowadays taking key relevance to 
overcome the sensitivity 
limits of the current generation of gamma--ray telescopes which see the sky through 
mechanical telescopes(see, e.g.,
Ref.~\citenum{Frontera97}) or coded mask systems (see, e.g.,
Ref.~\citenum{Ubertini03}).

Low energy focusing telescopes up to $\sim$ 70--80 keV have been successfully built (e.g. 
ASTRO-H~[\citenum{Kunieda10}] and NuSTAR~[\citenum{Hailey10}]), but beyond 100 keV an efficient 
way to focus photons appears to be the use of diffraction techniques from crystals.
The use of gamma--ray focusing optics is also crucial for improving the angular resolution, the 
best being now obtained with coded mask telescopes (about $\sim 15$ arcmin of the mask 
in the case of INTEGRAL/ISGRI).

\section{Open issues that can be settled only with deep soft $\gamma$-ray observations}

The astrophysical issues that are expected to be solved only with focusing  telescopes that cover the 
hard X--ray band from 80  to 511 keV and beyond are manifold. A discussion of them is extensively 
given in Ref.~[\citenum{Frontera11}]. Here we summarize some of them.

Soft Gamma Ray Repeaters (SGRs) and Anomalous X-ray Pulsars (AXPs) have raised many questions related to the
role of their strong magnetic field in the high energy emission. Furthermore, it is still not 
clear the belonging of these two types of source to the same class. The origin of their high energy 
component ($>$100~keV) is still not understood. 
To clarify the emission above this limit we need a better sensitivity than that of the current 
instruments at E $>$100~keV.

High energy emission mechanisms in compact galactic objects and  AGNs is also still not well understood.
The emission region can be investigated measuring the high energy cutoff and its relation to
the power-law photon index of the energy spectrum. Much more sensitive observations are needed, 
for both AGNs and compact galactic sources. 

%
Furthermore, AGNs physics is in the focus of astrophysicists also to establish the CXB origin. 
Most models assume a combination of unobscured, Compton thin and Compton thick 
radio-quiet AGN populations with different photon index distributions and fixed high energy
spectral cutoff (E$_c$). This assumption could be rejected or confirmed, with a large sampling
and well studied (in terms of sensitivity) AGNs. 

Positron production occurs in a variety of cosmic explosion and acceleration sites, and the observation 
of the characteristic 511 keV annihilation line provides a powerful tool to probe plasma 
composition, temperature, density and ionization degree. Compact objects - both galactic and 
extragalactic - are believed to release a significant number 
of positrons, leading to 511 keV gamma-ray line emission in the inevitable process of annihilation. 
A recent SPI/INTEGRAL all-sky map of galactic e$^-$/e$^+$ annihilation radiation 
shows an asymmetric distribution of 511 keV emission arond the Galactic Center that has 
been interpreted as a signature of 
low mass X-ray binaries with strong emission at photon energies $>$20 keV (hard LMXBs). 
Much more sensitive observations are needed to study the 
annihilation line origin, sources and their nature.

A gamma--ray telescope with a passband from 800 to 900 keV
could study on fascinating class of events: the explosion of Type Ia supernovae (SNe Ia). 
These explosions are the major contributors to the production of heavy elements. Hence they 
are a critical 
component for the understanding of the matter life cycle in the Universe and of the chemical 
evolution of 
galaxies. Because Laue lens telescopes allow the direct observation of radioactive isotopes 
that power 
the observable light curves and spectra, gamma-ray observations of SNe Ia can be performed with this 
type of instrument to get a breakthrough on the detailed physical understanding of 
SNe Ia. This is important for its own sake, but it is also necessary to constrain systematic
errors when using 
high-z SNe Ia to determine cosmological parameters. 
A sensitivity of $10^{-6}$~photons cm$^{-2}$~s$^{-1}$ to 
broaden gamma-ray lines allows observations of supernovae up to distances of 50-100 Mpc. 
Within this 
distance it is expected that there will always be a type Ia SN in the phase of gamma-ray 
line emission, 
starting shortly after explosion and lasting several months.

%
%

\section{The LAUE project: a focusing lens for soft  gamma-rays} 
\label{sec:laue}

The main goal of the LAUE project is to develop an advanced technology for building a Laue lens with broad
energy band (70/100--600 keV) and long focal length (up to 100 m), for space astrophysics. 
The project also faces an aged and difficult issue: the development of crystals suitable for
a lens. The project is supported by ASI, the Italian Space Agency.

The adopted lens assembly technology is new and it is the result of the experience gained thus far. 
In the previous attempt (HAXTEL project [\citenum{Frontera08}]) most of the mechanical errors and 
uncertanties in the lens assembly were related to the use of pins directly glued on each crystal, 
as a reference for the X-ray beam direction.

The gluing of pins introduces uncertainties and 
these  is summed to the second phase of the process, of counter-mask positioning and gluing 
phase on the carbon fiber support.

The new adopted technology consists in the positioning of the crystal tiles on the lens frame
under the control of a gamma--ray beam. Each lens crystal is correctly oriented when it focuses the 
beam photons in the lens focal plane. This position will be kept steady by 
gluing each crystal upon the lens frame. The lens frame is kept fixed during the lens assembling.

This method would allow to minimize uncertainty effects in orienting the crystal tiles in the lens and
would increase the crystal assembling rate in the lens with respect to that of the technology adopted so 
far [\citenum{Frontera08}]. The main contractor of the LAUE project, DTM Technologies (Modena), 
is responsible for the mechanical device (a robot) able to handle and set each crystal tile with the 
correct orientation for the diffraction. 

The lens is assumed to be made of a number of petals. One of these petal will be developed as a result of the 
LAUE project. The crystal production task for the petal is shared by two partners of the project: the Sensor and 
Semiconductor Laboratory (LSS) of the University of Ferrara and the CNR/IMEM, Parma. 

\begin{figure}[!t]
\begin{center}
\includegraphics[scale=0.8]{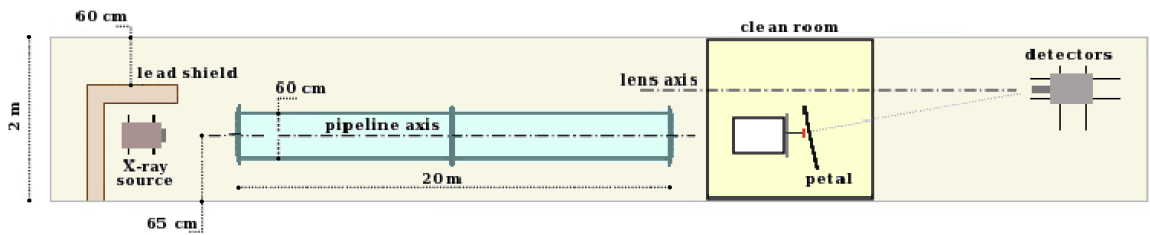}
\caption{\footnotesize Layout of the LARIX facility in which the petal will be built and tested.}
\label{fig:tunnel} 
\end{center}
\end{figure}

The lens assembly  apparatus is installed in the LArge Italian X-ray Facility (LARIX)
located in a tunnel (see Fig. \ref{fig:tunnel}) of the Physics Department of the University of Ferrara. 
For the lens assembling, it makes use of an X-ray generator with a fine focus of 0.2 mm 
radius with a maximum voltage of 320 kV and a maximum power of $\sim$ 1800 W. 
The photons coming from the X-ray tube are first collimated, then they pass through a 
20 m long under--vacuum pipeline. The X--ray entrance and exit windows of the pipeline are made
of carbon fiber 3 mm thick. The final beam collimation is performed at the exit window of the 
pipeline, in a clean room (class 10$^5$, US FED STD 209E Cleanroom Standards) in which the crystal 
assembly apparatus is located. The clean room is furthermore endowed with a thermal control 
(within $1\,^{\circ}{\rm C}$ accuracy) and an hygrometric control (relative humidity $\Phi$ = 60\% within 
an error of 10\%).

The final collimator aperture can be remotely adjusted in two orthogonal directions for 
divergence control. The beam through the final collimator is used to establish the 
crystal orientation in the lens.

The control of the single crystal focusing point is performed by means of two focal plane detectors: 
an X-ray imaging detector with spatial resolution of 300 $\mu$m and a cooled HPGe spectrometer 
with a 200 eV spectral resolution. Both are located on a rail and can be moved back and forth 
along the beam axis. Due to the maximum voltage of the available X--ray generator, the passband of 
the lens petal that will be produced is 80--300 keV.

\subsection{Crystals adopted for the lens petal} 

After a development phase, for the lens petal that will be built with
bent crystal tiles of Germanium (220), Silicon (111) and Gallium Arsenide (220) will be used.

The crystal cross section has been chosen to be 30 $\times$ 10 mm$^{2}$, with the longer side 
radially placed on the lens frame. The main advantages of the rectangular shape, 
together with the radial dispositon, concerns the focusing effect provided by bent crystals, 
which only acts in the radial direction. In such a way, a shorter tangential dimension provides 
a smaller defocusing factor, being proportional to the tile size.  
On the other hand, a bigger radial dimension makes the total number of crystal smaller, reducing
the error budget potentially caused by each crystal misalignment contribution. 

The thickness $t$ of the crystal tiles, for each type of crystal material, is a compromise 
between the need of a high effective area and the current limitation in the thickness imposed
by the current status of the bending technology adopted.
Moreover, the focal spot dimension is also linearly proportional to the thickness and an 
increase of this parameter causes a spreading of the PSF (for a 20 m focal length the longitudinal 
dimension of the PSF for a single crystal is roughly proportional to $t$/4).   
 
Bent crystals can be obtained with different methods [\citenum{Smither05}]. 
For space applications, a bent crystal can be obtained by 
growing a two component crystal (e.g., Si and Ge) with a concentration of each material
that changes along the growth axis. This method 
produces an intrinsic cuvature avoiding an external stress, even if the manufacturing turns out to be
difficult and awkward.

It has been recently demonstrated that indentations on the surface normal to the diffraction 
planes bend the crystal [\citenum{Bellucci11}]. 
For the LAUE project, an excellent cilindrical or spherical shape with the desired curvature radius
can be obtained, by finely 
tuning the parameters of the process (grooves number, width and depth of indentation, 
speed of the process). The bending has been successfully applied to Silicon and Germanium.
Following the theory for bent crystals [\citenum{Malgrange02,Keitel99}], the reflectivity of bent crystals
can be evaluated.

\section{Configuration of the petal that will be built and its expected performance}

\begin{figure}
\begin{center}
\includegraphics[scale=0.45]{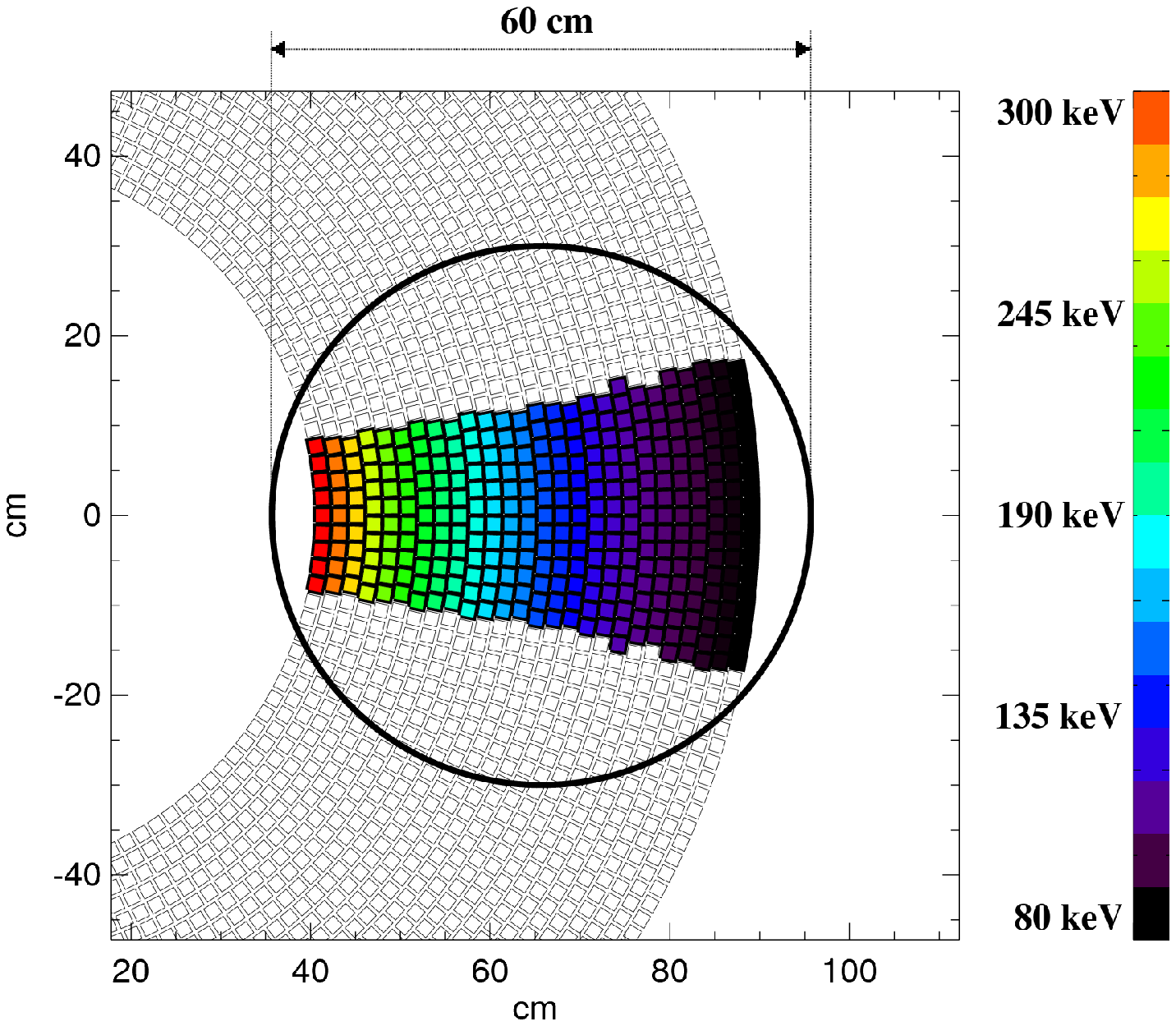}
\includegraphics[scale=0.45]{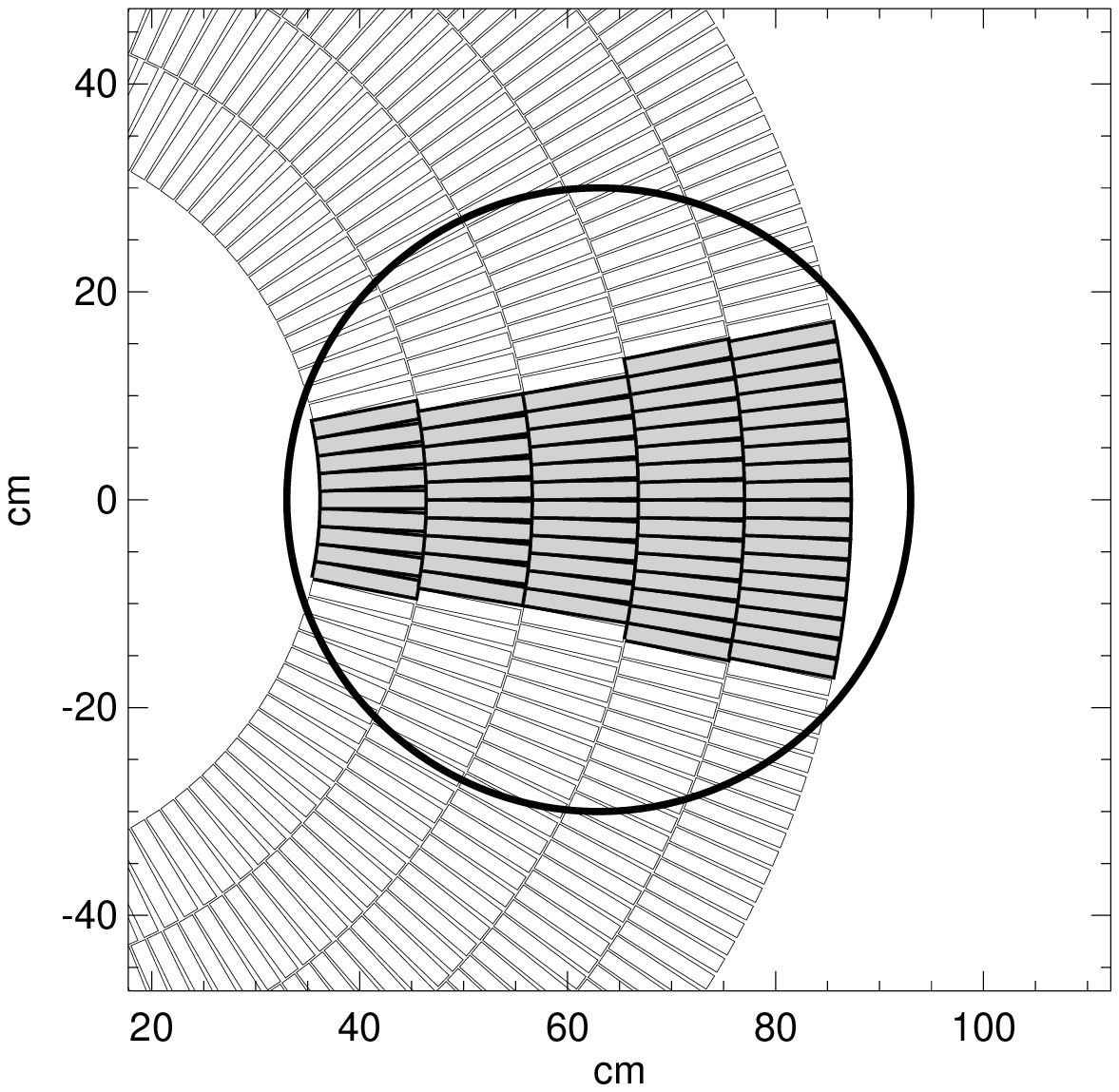}
\caption{\footnotesize Sketch of the petal extracted from an entire Laue lens, that will be 
built and tested in the Ferrara LARIX facility. {\em Left panel}: Configuration in the assumption
of a flat 1.5 $\times$ 1.5 cm$^2$ crystal cross section. Different colors give the  different energies
that will be reflected at 1st order diffraction. The energy scale of colors is also shown.
The black circle shows the size of the beamline within the petal has to be inscribed. 
{\em Right panel}: The petal configuration in the case of 
rectangular tiles (10 $\times$ 1.5 cm$^2$) that increases the filling factor and decreases the required number 
of tiles.}
\label{fig:petal} 
\end{center}
\end{figure}

In Table~\ref{tab:configuration} the main properties of the petal have been reported.
The petal is derived assuming a spherical lens equally divided into 20 petals.
The energy passband is defined by the inner and outer radius of the lens. For the
petal that will built, these values (see Table~\ref{tab:configuration}) are those allowed by the 
pipeline diameter within which the petal has to be inscribed to assemble the lens  petal
(see Fig. \ref{fig:petal}). 

The best distribution crystal materials within each ring can performed by using a genetic code
that distributes each crystal position of established cross section, available on each ring,
to the crystal material that satisfies pre-defined criteria (e.g. maximization of the 
effective area and/or maximum allowed derivative of the effecttive area with photon energy, etc).
%

\begin{table}[!h]
\begin{center}
\begin{tabular}{|l|l|}
\hline
Materials  and selected planes          &   Si(111), Ge(220), GaAs(220) \\
\hline
Energy passband (keV)                    &   80--300 keV \\
\hline
Focal length (m)       	                 &   20 \\
\hline
Petal inner/outer radius         	 &   41/93 cm\\
\hline
Crystal cross section (mm$^2$)           &   30 x 10 \\
\hline
Number of crystals for the petal         &   $\sim$350\\
\hline
Number of crystal rings                  &   18 \\
\hline
\end{tabular}
\end{center}
\caption{\footnotesize Main properties of the petal that will be built with the LAUE project.}
\label{tab:configuration}
\end{table}

\begin{figure}[!h]
\begin{center}
\includegraphics[scale=0.45]{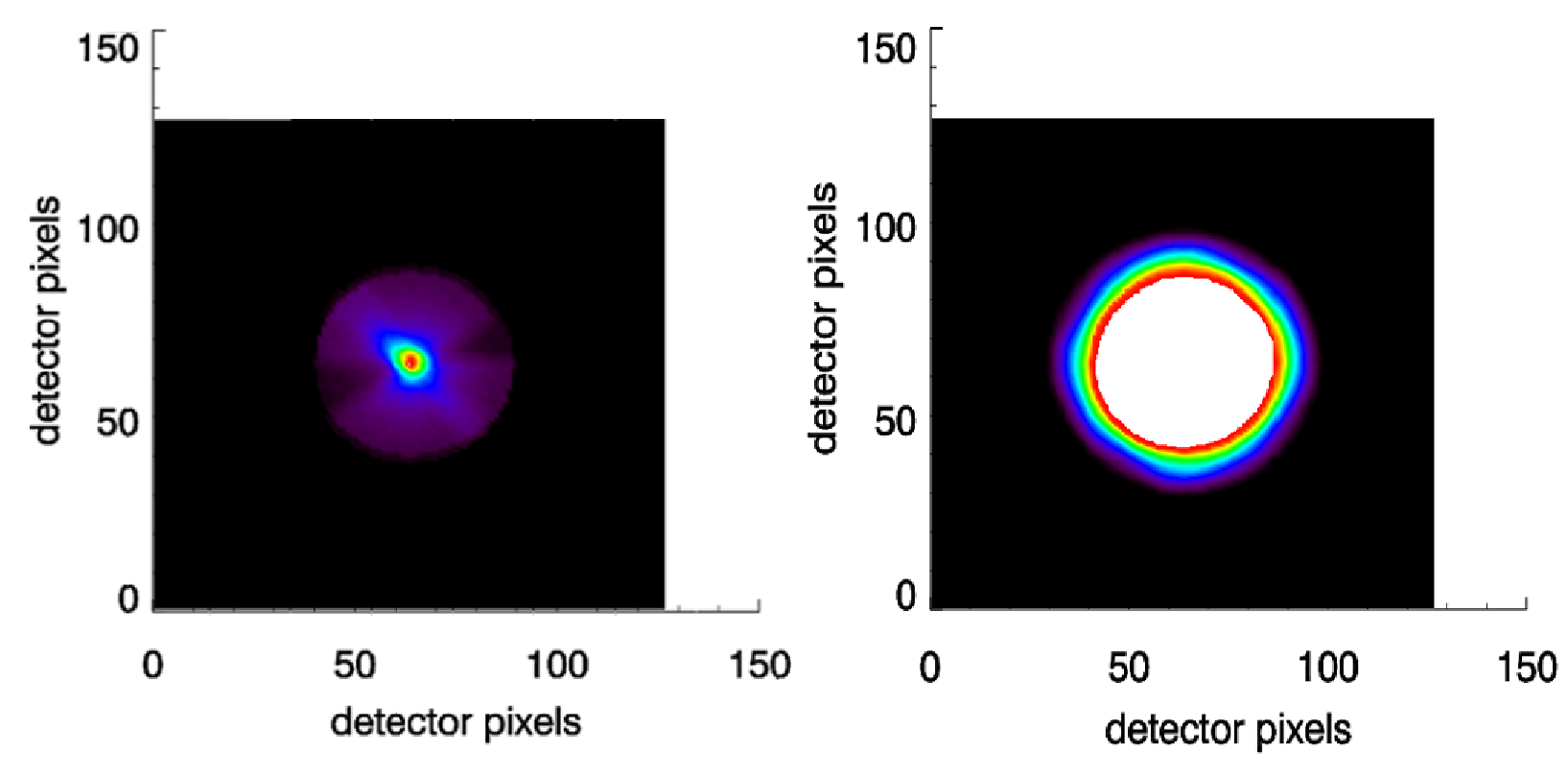}
\includegraphics[scale=0.34]{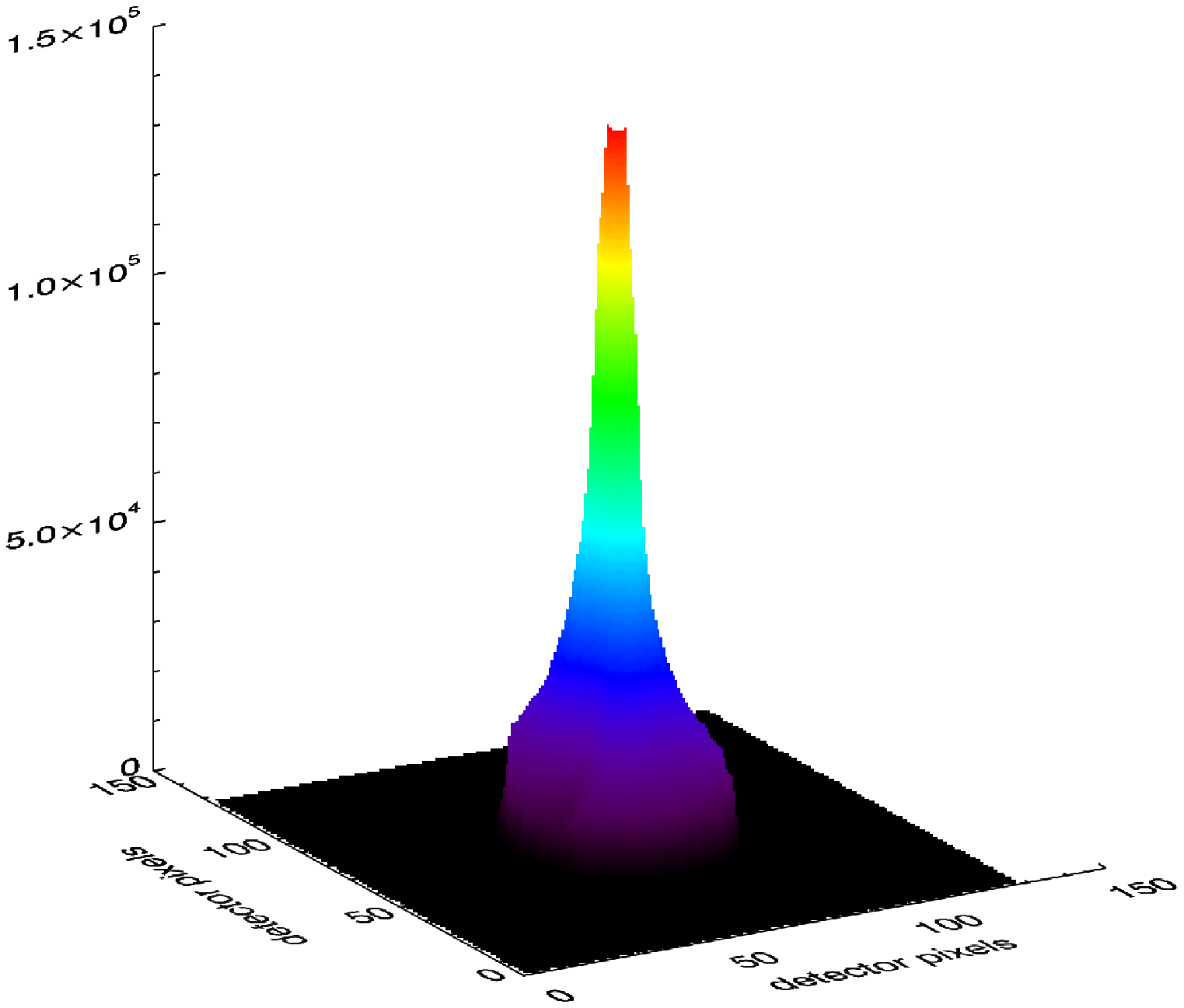}
\includegraphics[scale=0.34]{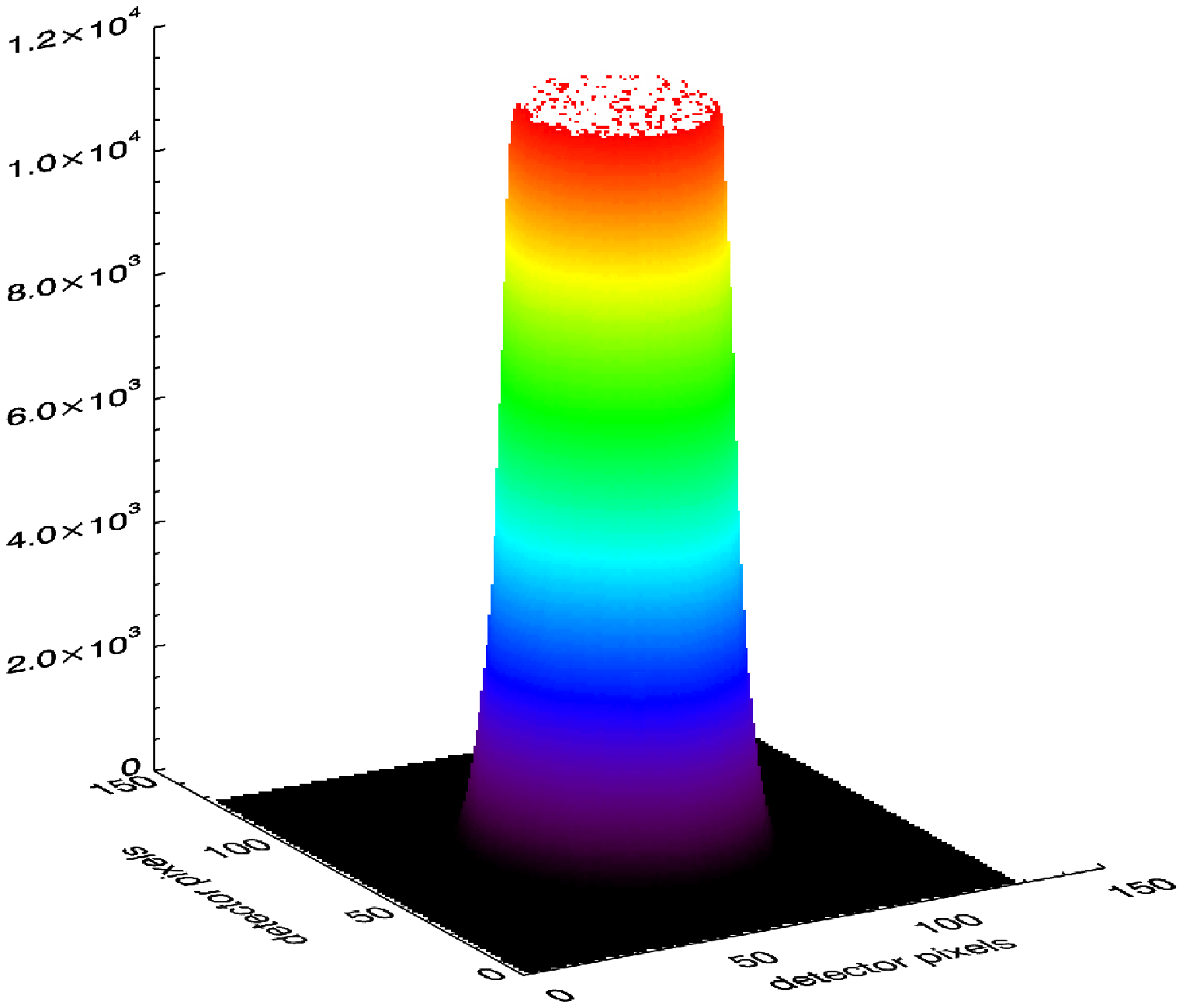}
\caption{\footnotesize Simulated PSF for a ring made of curved crystals ({\em left panel}) 
and flat mosaic crystals ({\em right panel}).}
\label{fig:spot.mosaic.bent.TOP.and.3D}
\end{center}
\end{figure}

The expected PSF of the assumed lens, in both cases of flat mosaic crystals and bent crystals,
is shown in Fig.~\ref{fig:spot.mosaic.bent.TOP.and.3D}. The PSFs have been obtained with a Monte Carlo
code assuming a single ring made either with Silicon bent crystals or flat mosaic crystals, respectively.
Crystals are simulated to be perfectly aligned with respect to the theoretical diffraction angle.
The same cross section of 1.5 $\times$ 1.5 cm$^2$ is assumed the same for both types of crystals 
which are considered to focus in a narrow band centred at 150 keV. Crystal orientation in the lens is 
that required by the Bragg diffraction.

\newpage

As can be seen, the three-dimensional plots show that in the case of bent crystals 
the PSF is significantly narrower than that obtained in the case of flat mosaic crystals. The 
number of photons collected in the center of the focal spot is more than one order of magnitude higher in 
case of bent crystals.

\begin{figure}
\begin{center}
\includegraphics[scale=0.25]{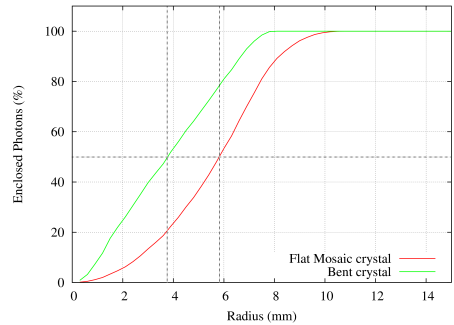}
\caption{\footnotesize Radial profile of the enclosed photons for  perfectly 
oriented  crystals in the lens for both cases of bent crystals and flat mosaic crystals.}
\label{fig:enclosed.bent.mosaic}
\end{center}
\end{figure}

The difference between the two types of crystals (bent or flat mosaic)  is 
better shown if the percentage of enclosed photons as a function of the distance from the focus
is plotted (see Fig. \ref{fig:enclosed.bent.mosaic}).

The better focusing capability of bent crystals is
evident, for a bent crystal the 50\% of the total photons are enclosed at a radius of 3.75 mm, 
the radius at the same percentage is 5.82 for a flat mosaic crystal.

\begin{figure}
\begin{center}
\includegraphics[scale=0.4]{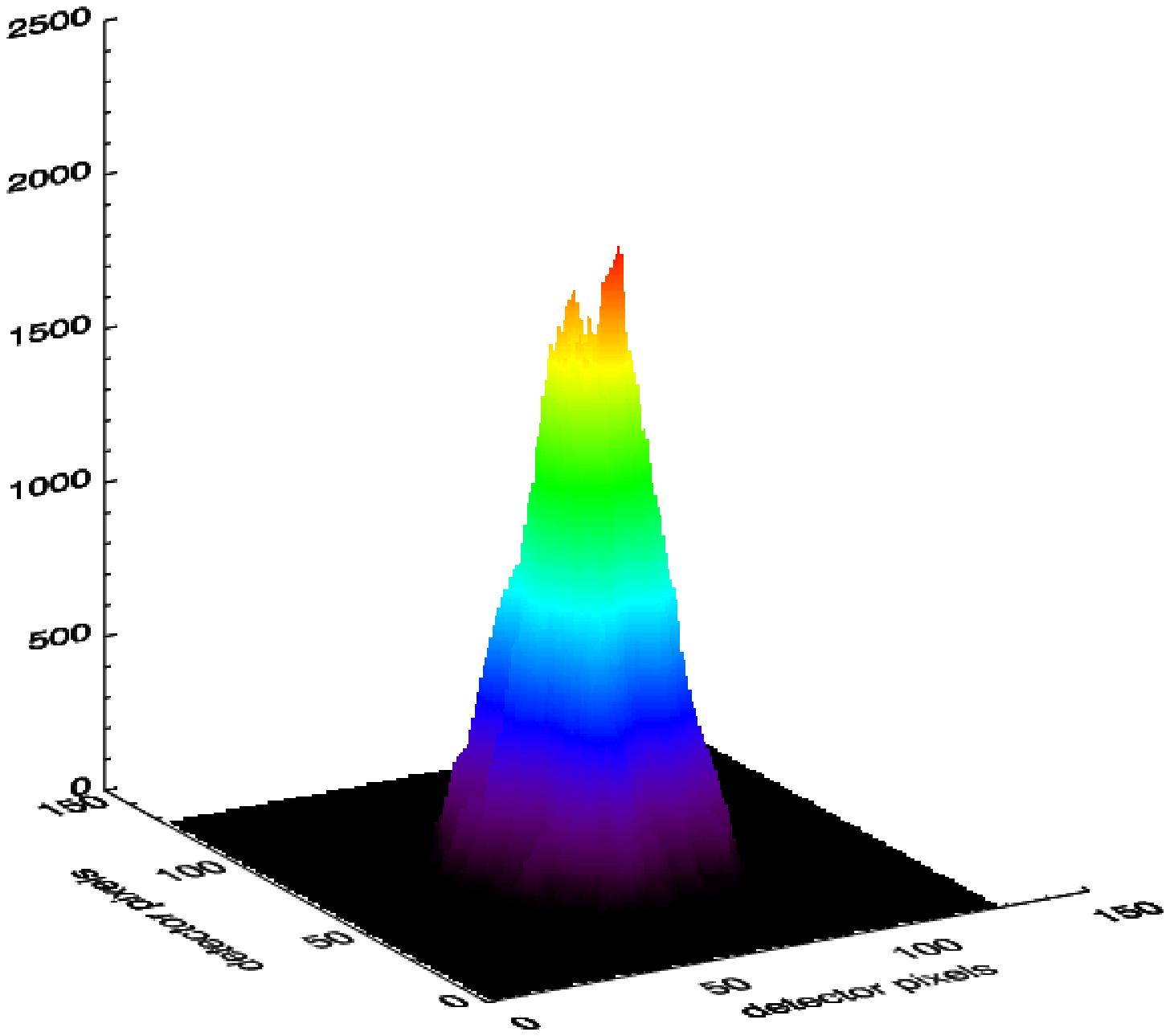}
\includegraphics[scale=0.4]{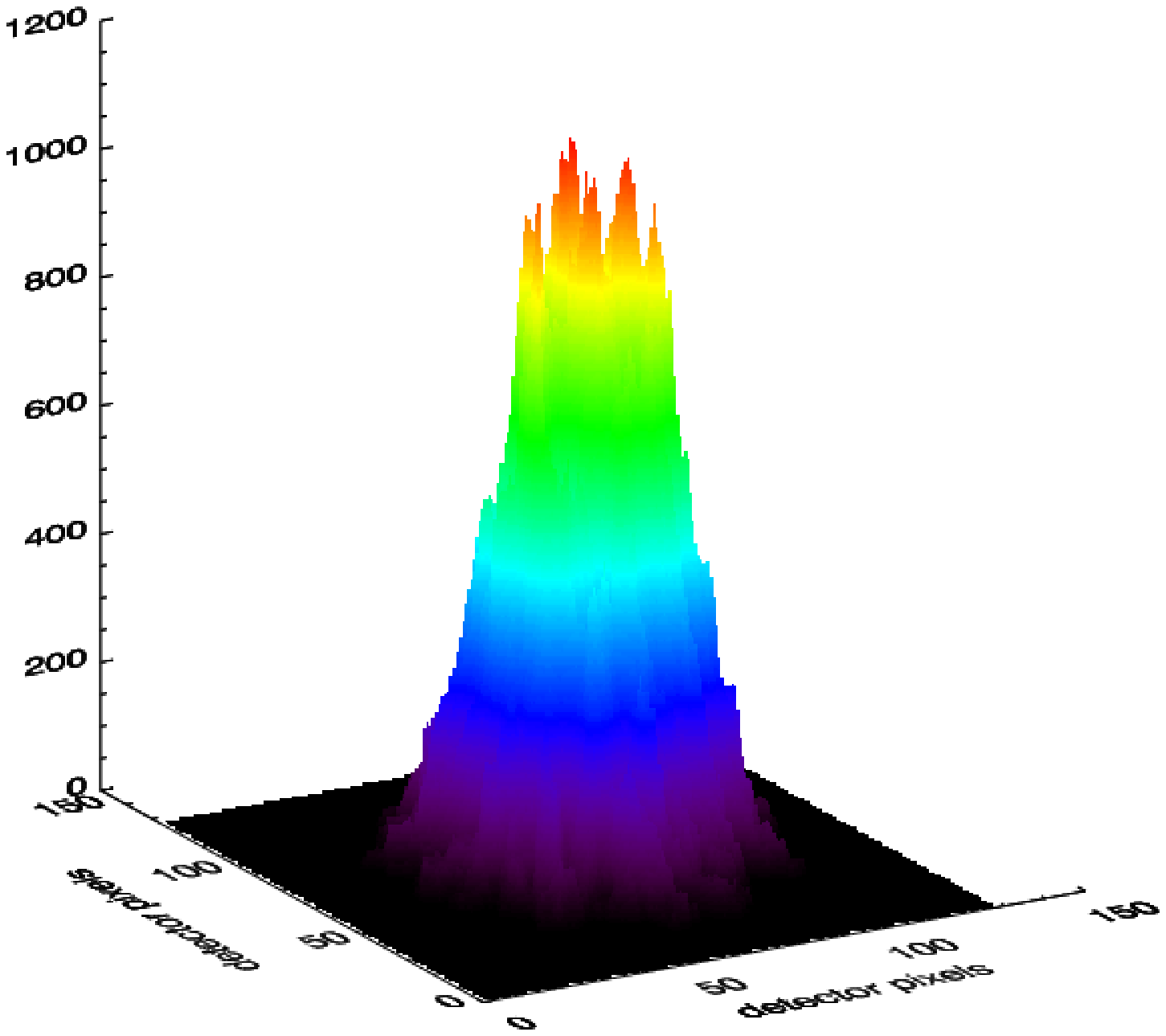}
\caption{\footnotesize Simulated PSF for a ring made of curved crystals with a misalignment 
of 0.5 arcmin (left) and 1 arcmin (right).}
\label{fig:PSFmisaligned.bent}
\end{center} 
\end{figure}

With the same Monte Carlo code, the effect of a misalignment has been investigated for bent crystals (see 
Fig. \ref{fig:PSFmisaligned.bent}) in which each tile has been simulated to be affected by misalignment angles 
with gaussian distribution around the angle 
of perfect  position on the lens, with a spread of 30 arcsec ($left~panel$) and 60 arcsec ($right~panel$).
In both cases the effect on the number of collected photons at peak drastically decrease of a factor 75 
for the former case, and of a factor 125 for the latter, with respect to the peak counts collected in the 
perfect aligned case. 

\begin{figure}
\begin{center}
\includegraphics[scale=0.25]{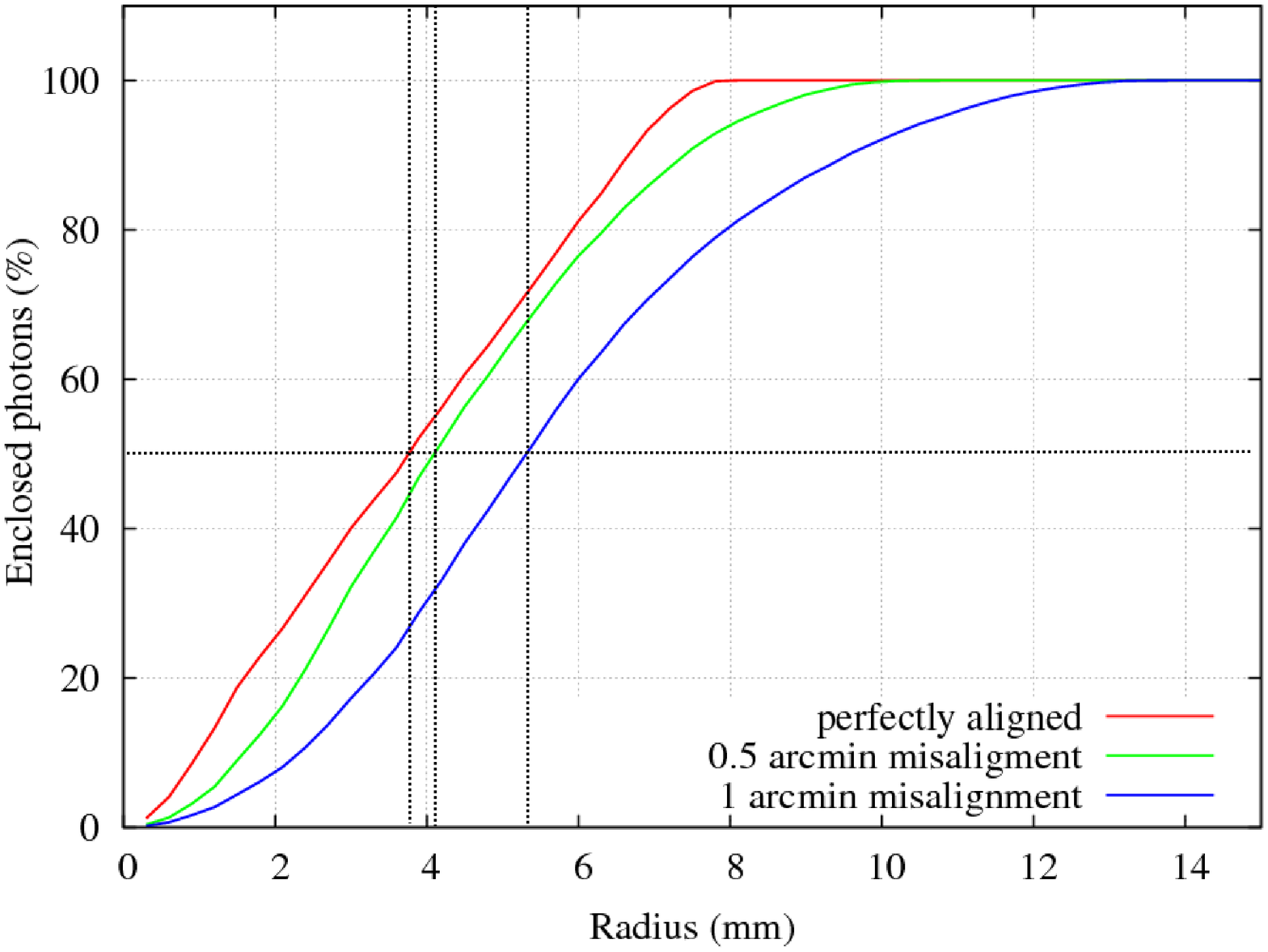}
\includegraphics[scale=0.25]{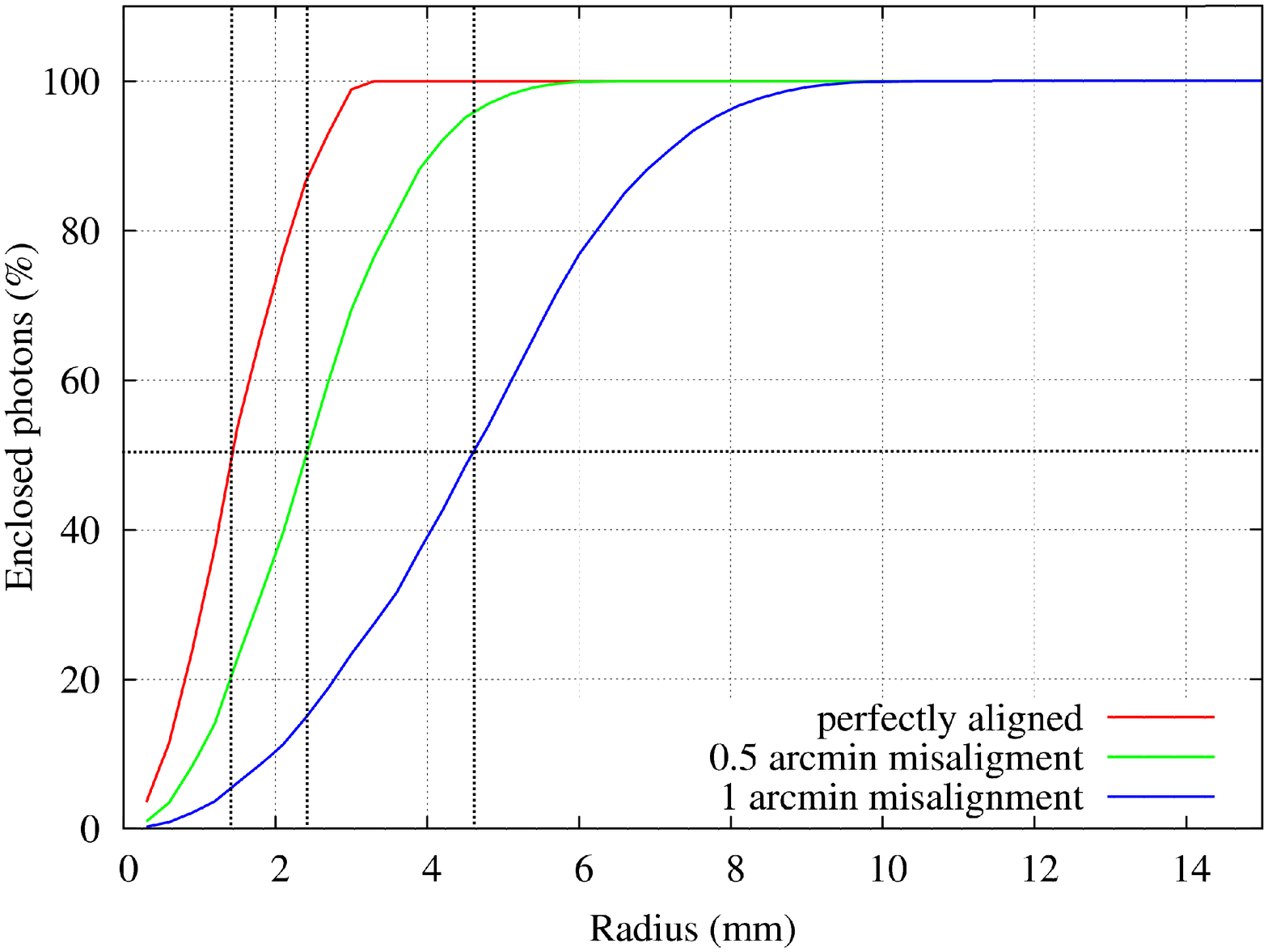}
\caption{\footnotesize Cumulative distribution curves of the focused photons in the
 case of  bent crystals for a perfect alignment, 
Gaussian misalignment of crystals with a $\sigma$ of 30 arcsec and 60 arcsec (fwhm)
 with respect to the orientation required by the Bragg diffraction.
{\em Left panel}: tile cross section of 1.5 $\times$ 1.5 cm$^2$. {\em Right panel}: 
tile cross section of 2 $\times$ 0.5 cm$^2$.}
\label{fig:cumulative.aligned.misaligned}
\end{center}
\end{figure}

The effect also strongly depends on the crystal cross section adopted 
(see Fig.~\ref{fig:cumulative.aligned.misaligned}). These results show the importance 
of a negligible misalignments in the crystal orientation. We expect,
from the adopted lens assembly technique a crystal misalignments not larger than 10 arcsec.

\newpage

\section{Conclusions}

From the already gained experience on Laue lenses (see, e.g., \citenum{Frontera08,Virgilli11}), we have started
a new project, LAUE, supported by the Italian Space Agency (ASI), devoted to the development of
an advanced lens assembling technology that we expect it eventually will allow to accurately build Laue
lenses for space astrophysics. The expected accuracy in the lens assembling would allow to build
lenses even with very long focal lengths (up to 100 m), a goal never achieved so far.
In addition to the develoment of the lens assembling technology the LAUE project is facing the 
crystal production of proper crystals for Laue lenses. Bent crystals appear the most suitable.

As a demonstration of the validity of the adopted technology, within the LAUE project, in the
LARIX facility of the University of Ferrara, we will build a lens petal of 20 m focal length.
Results of the petal lens built will be reported in one of the next SPIE conferences.

\acknowledgments     

The LAUE project is the result of big efforts made by a large number of organizations and people.
We would like to thank all of them, as the success of the petal assembling means a step forward on the
building of the whole lens. We aknowledge the ASI Italian Space Agency for its support to the LAUE project.


\bibliography{lauepaperFF}
\bibliographystyle{spiebib}

\end{document}